\documentclass[epj]{svjour}
\usepackage{graphics}
\usepackage{notoccite}
\usepackage[table,xcdraw]{xcolor}
\usepackage{amsmath}
\begin{document}
\title{Charmed and bottomed hadronic cross sections from a statistical model}
\author{Gábor Balassa\inst{1} \and György Wolf \inst{1} 
}                     
%
%
\institute{\inst{1}Institute for Particle and Nuclear Physics, Wigner Research Centre for Physics, H-1525 Budapest, Hungary}
\date{Received: date / Revised version: date}
%
\abstract{
In this work, we extended our statistical model with charmed and bottomed hadrons, and fit the quark creational probabilities for the heavy quarks, using low energy inclusive charmonium and bottomonium data. With the finalized fit for all the relevant types of quarks (up, down, strange, charm, bottom) at the energy range from a few GeV up to a few tens of GeV's, the model is now considered complete. Some examples are also given for proton-proton, pion-proton, and proton-antiproton collisions with charmonium, bottomonium, and open charm hadrons in the final state. 
\PACS{
      {PACS-key}{discribing text of that key}   \and
      {PACS-key}{discribing text of that key}
     } 
} 
\maketitle
\section{Introduction}
\label{intro}
Low and medium energy (up to 20-30 GeV) inclusive and exclusive cross sections are important inputs for transport calculations in heavy ion collisions \cite{1,2,3,4,5}. Most of these cross sections are coming from measurements \cite{6,7}, and/or from some effective model calculations \cite{8} e.g. Walecka type models \cite{9}. For many interesting processes, measurements are nonexistent, or very sparse e.g. for strange vector mesons $K^*(892)$ \cite{10,11,12} or charmonium cross sections near threshold \cite{13,14}. As some of these hardly accessible processes could be important in the investigation of strongly interacting nuclear matter formed in heavy ion collisions \cite{15,16,17}, phenomenological model calculations or extrapolation from fits to available data are necessary. In \cite{18} we proposed a model, based on the statistical Bootstrap approach \cite{19,20,21}, which were able to give reliable estimates to low/medium energy cross sections (few GeV) for exclusive hadronic processes. The model is extended in \cite{22} to be able to give estimatations to inclusive processes as well, with an approximated error bound coming from the model uncertainties. Using the extended model, we were able to fit the charm quark creational probability to existing charmonium data. 
Using the fitted value, the model had been proved to be able to describe inclusive charmonium production in proton-proton, and in pion-proton collisions. In this paper, we further extend the model by introducing more particles into the calculations, and by fitting the bottom quark creational probability. These extensions will slightly change the previously fitted value for the charm quark creational probability as well. 
Including heavy quarks into statistical models have been done in the past \cite{60}, where usually a suppression factor is introduced into the quarks momentum distribution function, which measures the deviation from chemical equilibrium. In our model the quark creational probabilities are taking over this role, which were fitted to measured cross sections, with final states containing heavy quarks.
With the modifications the model is now considered complete, as the energy range, we intend to use is far below the top quark mass, so its inclusion in the model is not necessary. The paper is distributed as follows. In Sec.\ref{sec:1} a basic formulation is given, which covers the main ingredients of the model. In Sec.\ref{sec:2} the fitting procedure for the charm and bottom quark creational probabilities is covered, while in Sec.\ref{sec:3} some charmonium, bottomonium and open charm inclusive cross sections are calculated for proton-proton, proton-antiproton and pion-proton collisions. After these examples Sec.\ref{sec:4} concludes the paper.

\section{Basic formulation}
\label{sec:1}
The model is based on the assumption that during a collision process a so called fireball is formed, which after a short time, will hadronize into a specific final state. The collision process therefore can be factorized into an initial dynamical part forming the fireball, and a mixed dynamical and statistical part, which describes the hadronization, as it is shown in Eq.\ref{eq:1}
\begin{eqnarray}
\label{eq:1}
\nonumber
&{\displaystyle \sigma^{n \rightarrow k}(E)=\Bigg( \int \prod_{i=1}^{n} d^3p_i R(E,p_1,...,p_n) \Bigg) } \\
&{\displaystyle \times \Bigg( \int \prod_{i=1}^{k} d^3q_i w(E,q_1,...,q_k) \Bigg)}
\end{eqnarray}
where $ \sigma^{n \rightarrow k}$ is a generalized transitional probability for an $n \rightarrow k$ process, $p_i$ is the 3-momenta of the incoming particles, $E$ is the center of momentum energy of the $n$-body collision, and $q_i$ is the 3-momenta of the outgoing particles. The function $R(E,p_1,...p_n)$ describes the initial stage of the collision, while $w(E,q_1,...,q_k)$ describes the hadronization of the fireball. Throughout this paper, we only consider two-body collisions, so $\sigma^{2 \rightarrow k}$ becomes the cross section of the collision. It is assumed that the dynamical part is described by the inclusive cross section of the colliding particles, so the integral in the first bracket simply becomes the inelastic cross section of the two-body reaction. This assumption is widely used in many statistical models in the past \cite{23,24}. Although it is worth to mention that it is possible to include more than two-body reactions as well, where only the initial dynamical part $R(E,p_1,...,p_n)$ has to be changed accordingly to some theory. The integral in the second bracket describes the fireball hadronization, and will be denoted by $W(E)$. The hadronization probability depends on the following factors: (1) Fireball decay scheme probability (2) Fireball density of states (3) Phase space of the final state (4) Quark combinatorial factor (5) Quark creational probabilities. In the followings a short description is given to each of the ingredients. 

After the fireball is formed, there is a possibility that the fireball with invariant mass $M$ hadronizes into some final state or decays into smaller fireballs with invariant masses $M_1,M_2$. Each of these smaller fireballs are then again choose to hadronize or decay into smaller fireballs. At the end of the chain there will be a specific number of fireballs, with invariant masses $M_1,M_2,…M_k$, and each of them ultimately decay into a specific hadronic final state. The probability of the number of fireballs is given by kinematical considerations and is described in detail in \cite{18}. The hadronization procedure is based on the statistical bootstrap approach, which sets the probability for the number of hadrons coming from one fireball to $P_2^{fb} = 0.69$ and $P_3^{fb}=0.24$ \cite{25}. More particles have negligible probabilities compared to the two-, and three body cases, therefore in our model the possible number of hadrons coming from one fireball is two, or three. However it is straightforward that due to the many number of smaller fireballs the number of hadrons in the full final state could be much more than three.

Following the bootstrap approach the hadronization into final state particles should depend on the density of states (DOS), and also the two-, and three body phase spaces with a slight difference on how we handle non-stable and stable particles. The density of states is given by:
\begin{equation}
\label{eq:2}
\rho(E) = \frac{a\sqrt{E}}{(E_0+E)^{3.5}}e^{\frac{E}{T_0}},
\end{equation}
where $T_0$ is the hadronization temperature, $E_0=500$ MeV is a cut off parameter, and $a$ is an irrelevant normalization factor, which disappears in the final results. These parameters are fitted using experimentally measured particle multiplicities and momentum spectra for protons, pions and kaons in \cite{26}, and in \cite{19}. It is also mentioned that $T_0$ could be essentially anything from $130$ MeV to $170$ MeV (with some corresponding changes in $E_0$ and $a$), giving almost the same mass spectrum, therefore, as our model is not really sensitive to $E_0$ and $a$, we did our own fit to $T_0$, using exclusive cross section ratios in \cite{18}, giving $T_0=160$ MeV.

In the model, stable and non-stable hadrons are distinguished, which is manifested in the phase spaces considered. A hadron is considered a resonance, if it has a non-negligible width and is able to decay into a lower lying hadronic state via the strong interaction. Taking into consideration the two types of particles, the phase space is written as:
\begin{eqnarray}
\label{eq:3}
\nonumber
&{\displaystyle\Phi_k(M,m_1,..,m_k) = V^{k-1} \times} \\
& {\displaystyle\Big( \int \prod_{r \in R} dE_r \Big) \Big( \int \prod_{i=1}^k d^3 \mathbf{q}_i \Big) \prod_{r \in R}^{}F_r^{BR}  \times} \nonumber \\
& {\displaystyle \delta \Big( \sum_{j=1}^k E_j - M \Big) \delta \Big( \sum_{j=1}^k \mathbf{q}_j \Big)}
\end{eqnarray}
where $\mathbf{q}_j$ is the three momenta of the j'th particle, $E$ is the energy, $k=2,3$ the number of hadrons in the final state of a fireball with invariant mass $M$, $V$ is the interaction volume, set to a corresponding interaction radius of $r=0.5$ fm, and $F_r^{BR}$ is a Breit-Wigner factor given by:
\begin{equation}
\label{eq:4}
  F_r^{BR}= \frac{1}{\pi}\frac{M_r^2 \Gamma_i}{(M_r^2-m_r^2)^2+M_r^2 \Gamma_r^2},
\end{equation}
where $M_r$ is the invariant mass of the resonance, with a corresponding decay width of $\Gamma_r$, and a pole position $m_r$. The product $\prod_{r \in R}$ goes through all of the resonant particles, while for stable hadrons only the momentum integral remains, with $F_r^{BR}=1$.

The last ingredient to the hadronization probability is the quark combinatorial factor and the corresponding quark creational probabilities. This is, in some way, similiar to the parton model \cite{27,28}, where the partonic cross sections are integrated out with the parton density functions giving a hadronic cross section. In this model the number of quark-antiquark pairs are estimated from simple phase space considerations and is given by \cite{29}:
\begin{equation}
\label{eq:5}
N(E)=\frac{1+\sqrt{1+E^2/T_0^2}}{2},
\end{equation}
where $T_0$ is the hadronization temperature. For each quark there is a creational probability $P_i$, $i=u,d,s,c,b$ and the quark distribution function is then given by the multinomial distribution:
\begin{equation}
\label{eq:6}
F(N,n_i) = \frac{N(E)!}{\prod_{i=u,d,s,c}n_i!} \prod_{i=u,d,s,c} P_i^{n_i},
\end{equation}
where $n_i$ is the number of quarks of type $i=u,d,s,c,b$.

The distribution function gives the probability of a specific quark number configuration. In our model, we take the configuration, where the probability is at maximum, which corresponds to a quark number $\overline{n}_i=P_i N$. With the number of quarks, the number of possibilities of a final state is then calculated with simple combinatorics, and multiplied by a color factor given by the number of colorless combinations, then normalized by the possible two-, and three body final state factors, giving a final quark combinatorial probability. Finally this probability is multiplied by the maximum value of the distribution function, giving the final quark combinatorial factor $C_{Q_i}$. When the expected number of quarks are very small, which is the case for the heavy quarks with small quark creational probabilities, a suppression factor is introduced. This is described in Sec.\ref{sec:2}, where we fit the charm and bottom quark creational probabilities. 

If the calculated process respects the conservation laws and the quarks/antiquarks from the colliding particles are also included into the combinatorial factors, meaning $n_i = \overline{n}_i+n_{initial}$ will be the number of a specific quark/antiquark, there will be no quarks or antiquarks left unpaired at the end, and every quark, which is not used in the final state, could annihilate with a corresponding antiquark.

Putting together the previously described factors the normalized hadronization probability for $k$ fireball is given by Eq.\ref{eq:7}.
\begin{eqnarray}
\label{eq:7}
\nonumber
&{\displaystyle W_{k,i_1..i_k}(E) =P_k^{fb}(E)  \frac{1}{Z_k}\frac{1}{N_{i_1,..i_k}!}\int_{x_{min}}^{x_{max}} \prod_{a=1}^{k} } \Bigg[ dx_a  \times \\ 
& {\displaystyle\times \frac{T_{i_a}(x_a)}{\sum_j T_j(x_a)}  \delta \Big( \sum_{a=1}^k x_a-E \Big)  \Bigg]},
\end{eqnarray}
where $P_k^{fb}(E)$ is the fireball decay scheme probability for $k$-fireballs, $N_{i_1,...i_k}$ is a symmetry factor counting the fireballs with the same hadrons after hadronization, $x_{min}$ and $x_{max}$ limits are given by kinematical considerations, and $Z_k(E)$ is the energy-dependent k-fireball normalization factor given by:
\begin{eqnarray}
\label{eq:8}
&{\displaystyle Z_k(E)=  \sum_{<l_1..l_k> \in S} \frac{1}{N_{l_1,..l_k}!} \int_{x_{min}}^{x_{max}} \prod_{a=1}^{k} } \Bigg[ dx_a  \times \\ \nonumber
& {\displaystyle\times \frac{T_{l_a}(x_a)}{\sum_j T_j(x_a)}  \delta \Big( \sum_{a=1}^k x_a-E \Big)  \Bigg]},
\end{eqnarray}
where the subscript $<l_1..l_k> \in S$ means that the summation goes for all of the possible processes having quantum numbers $S$, where $S$ includes baryon number, electric charge, strangeness, charmeness, and bottomness. Also, for simplicity the $T_i(E)=C_{Q_i}(E)P_{n_i}^{H,i}(E)$ notation is introduced, where $C_{Q_i}(E)$ is the quark combinatorial factor, and $P_{n_i}^{H,i}(E)$ is given by:
\begin{equation}
\label{eq:9}
P_{n}^{H,i}(E) = P_n^d \frac{\Phi_n(E,m_1,..,m_n)}{\rho(E)(2\pi)^{3(n-1)} N_I!} \prod_{l=1}^n (2s_l+1),
\end{equation}
where $n=2,3$ the number of hadrons coming from one fireball, $s_l$ is the spin of the hadron, $P_n^d$ is the probability of two-, or three hadrons, and $N_I$ is a symmetry factor given by the number of indistinguishable hadrons in the final state of one fireball. In the normalization, only processes with the quantum numbers of the initial state is considered and for many fireballs only the full final state, with all the produced hadrons has to respect the conservation laws, and not each fireball separately.

\section{Quark creational probabilities for bottom and charm quarks}
\label{sec:2}
At high energies it is expected that the quark creational probabilities will be asymptotically the same, so $P_u=P_d=P_s=P_c=P_b$, with $P_u+P_d+P_s+P_c+P_b=1$, however in the energy range we are interested in (2-20 GeV) the equality is only valid for the up and down quarks. For the strange, charm and bottom quarks $P_s,P_c,P_b$ has to be fitted from experiments or has to be estimated from some underlying theory. We choose to fit the parameters from experiments, where for the strange quark, the exclusive $p \bar{p} \rightarrow K^+ K^-$ process was used, assuming a constant $P_s$ \cite{18}. For the up, down and strange quarks the values are $P_u=0.425$, $P_d=0.425$, and $P_s=0.15$. 
For the charm and bottom quarks the fit is not that straightforward due to their considerable masses and therefore their huge suppression. Due to the very small charmed and bottomed probabilities it is assumed that only one charm/bottom quark-antiquark pair is created, but the following arguments can be made to more charm and bottom quarks as well. Due to this assumption the constrained probability (one charm/bottom quark is certainly created) obtained from the quark number distribution Eq.\ref{eq:6}, will be much smaller than the non-constrained maximal probability. The suppression for both heavy quarks is coming from the ratio of the maximal and the constrained probability, obtained from the quark number distribution mass function, and can be expressed as:
\begin{equation}
\label{eq:10}
\gamma_{c,b} = \frac{F(N,n_u',n_d',n_s',n_c',n_b')}{F(N,n_u,n_d,n_s,n_c,n_b)} = \frac{n_s! P_{c,b}}{(n_s-1)! P_s}=P_{c,b} N
\end{equation}
where it is assumed that only one heavy quark is created, replacing one strange quark, meaning $n_{c,b}'=1$, and $n_s'=n_s-1$, while $n_u'=n_u$ and $n_d'=n_d$ . It is evident that the suppression is energy dependent even when the quark creational probability is constant and the ratio of the global and the constrained maximum tends to unity as the energy goes higher. After this point the global maximum value of the quark number distribution mass function is used and the expected number of heavy quarks, which corresponds to the global maximum are $n_{c,b}=P_{c,b} N$, so the transition between the constrained and the global maximum case is continous. 

\begin{figure}
\resizebox{0.5\textwidth}{!}{%
  \includegraphics{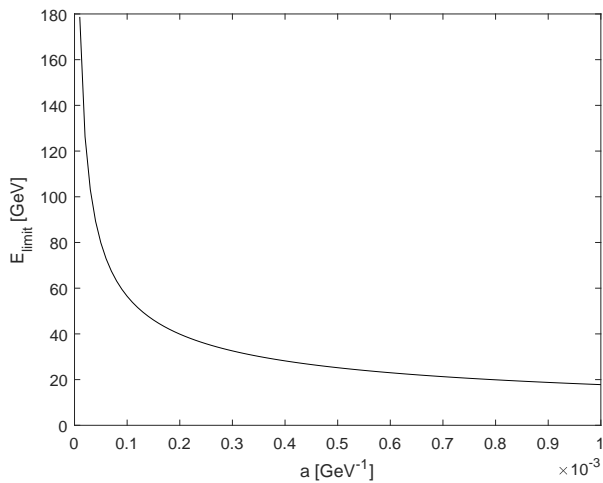}
}
\caption{Limit energy dependence on the slope parameter $a_{c,b}$ in the heavy quark creational probability $P_{c,b}=a_{c,b}E$. The limit energy marks the energy, where the constrained maximum of the quark number distribution function in Eq.\ref{eq:6} coincides with its global maximum.}
\label{fig:1}       
\end{figure}

A constant quark creational probability is however not too realistic, because at high energies these parameters are essentially have to be equal, therefore we assume a simple linear relationship between the energy and the quark creational probability $P_{c,b} = a_{c,b}E$, for the charm and bottom quarks. For the strange quarks, we keep the constant $P_s$ value, as it was sufficient to describe exclusive $K$, $\Lambda$, and inclusive strange vector meson production cross sections in the energy range of a few tens of GeV's, where we intend to use the model. It is possible that at higher energies these values will change, but at this relatively low energy range the linear relationship for the heavy quarks and the constant values for the up, down and strange quarks seem to be good approximations. It is also possible that the rise of the strange quark creational probability is so small that $P_s$ can be approximated by a constant value at the energy range considered. Using Eq.\ref{eq:10} and the energy dependent $P_{c,b}$, the heavy quark suppression can be expressed as:
\begin{equation}
\label{eq:11}
\gamma_{c,b} = a_{c,b} \frac{E+E \sqrt{1+E^2/T_0^2}}{2},
\end{equation}
where $a_{c,b}$ slope parameters have to be determined from experiments. The suppression is energy dependent through the number of quarks, and $P_{c,b}$. This is of course only valid until $\gamma_{c,b}$ reaches unity, as after that point the global maximum will naturally give one heavy quark, and it is not necessary to constraint the quark number distribution anymore. The actual energy when this happens, depends on the slope parameters $a_{c,b}$ of the charm and bottom quarks and can be calculated setting $\gamma_{c,b} = 1$ and solving the following equation:
\begin{equation}
\label{eq:12}
a_{c,b} \frac{E_L+E_L \sqrt{1+E_L^2/T_0^2}}{2} - 1 = 0,
\end{equation}
where $E_L$ is the limit energy. On Fig.\ref{fig:1} the limit energy is shown as the function of the slope parameter $a_{c,b}$.

As the quark creational probability depends only on $a_{c,b}$, this is the only parameter that has to be fitted from experiments. In \cite{22} a very simple fitting procedure is used where, $a_c$ was fitted only to one data point from the $\pi^- p \rightarrow J/\Psi X$ reaction and checked if it gives the correct values to different measurement points, in different collisions as well. In this paper a least squares fitting method is used, where an optimum point is sought in the error function, defined as:
\begin{equation}
\label{eq:13}
Error = \sqrt{\frac{1}{N}\sum_{i=1}^{N}\Big( \frac{\text{meas}_i-\text{model}_i}{\text{meas}_i} \Big)^2}
\end{equation}
where meas$_i$ is a measurement point, model$_i$ is the corresponding model calculation, and $N$ is the number of measurement points used for fitting. 

\begin{figure}
\resizebox{0.5\textwidth}{!}{%
  \includegraphics{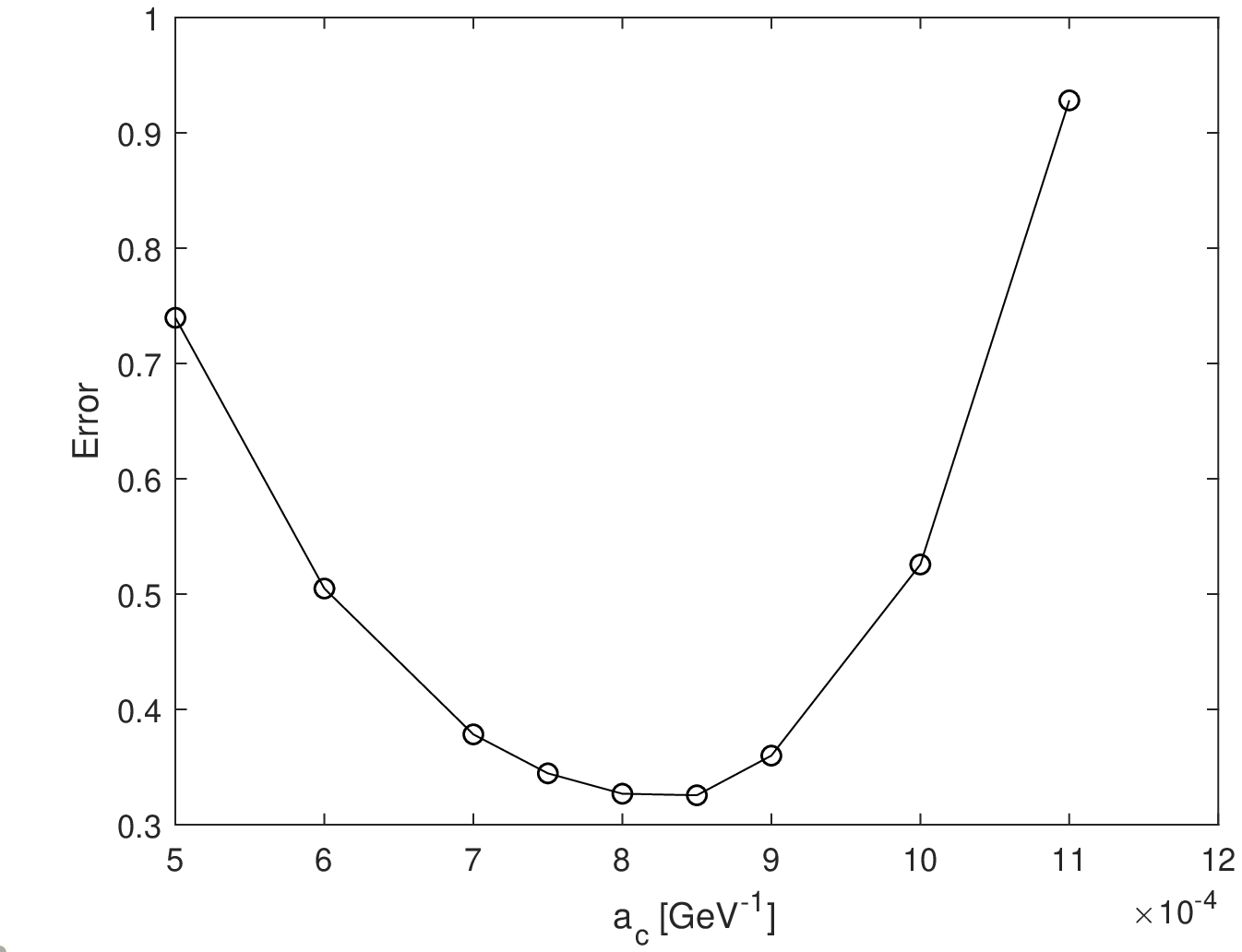}
}
\caption{Error function for the inclusive $J/\Psi$ production. The minimum corresponds to the slope $a_c = 8.5\cdot10^{-4}$ GeV$^{-1}$.}
\label{fig:2}       
\end{figure}

For the charm quarks the inclusive $\pi^- N \rightarrow J/\Psi X$, and $p p \rightarrow J/\Psi X$ reactions were used, where the measured values were collected from \cite{30,31,32,33,34,35,36,37,38,39,40,41}. To calculate the inclusive cross section from the model, apart from the prompt $J/\Psi$ production, the radiative and hadronic decays from the excited charmonium states $\Psi(3686)$, $\chi_{c1}$, and $\chi_{c2}$ are included as well, using a two fireball decay scheme. In Fig.\ref{fig:2} the calculated error for the $J/ \Psi$ production cross section is shown, where a clear minimum can be seen at $a_c=8.5 \cdot 10^{-4}$ GeV$^{-1}$, which is really close to our previous fit in \cite{22}. The cause of the difference is that now more particles are introduced in the model e.g. new charmed mesons/baryons, bottomed mesons/baryons, which in exchange slightly changes the normalization values as well. 

\begin{figure}
\resizebox{0.5\textwidth}{!}{%
  \includegraphics{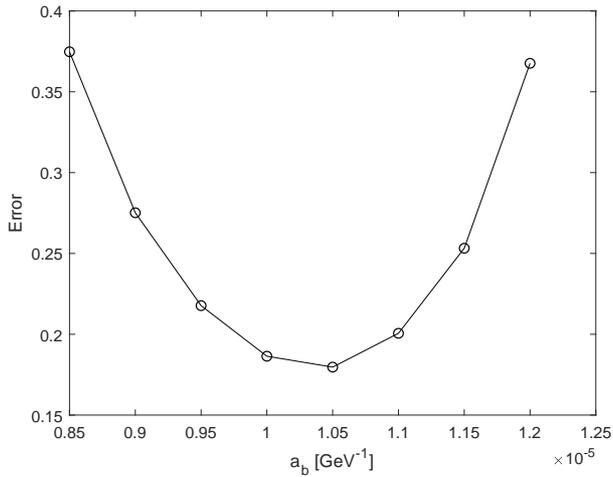}
}
\caption{Error function for the inclusive $\Upsilon(1S)$ production. The minimum corresponds to the slope $a_b = 1.05 \cdot 10^{-5}$ GeV$^{-1}$.}
\label{fig:4}       
\end{figure}

For the bottom quarks the fitting procedure is the same as it was for the charm quarks. The process $\pi^- p \rightarrow \Upsilon (1S) X$ is used for fitting, where the measured points are taken from \cite{42,43}. In the model calculations apart from the direct $\Upsilon (1S)$ meson, the excited onium states $\Upsilon (2S)$, $\Upsilon (3S)$, $\chi_{b1}(1P)$, $\chi_{b2}(1P)$, $\chi_{b1}(2P)$, and $\chi_{b2}(2P)$ are used as well. The results for the error function can be seen in Fig.\ref{fig:4}, where a minimum is obtained, now at $a_b = 1.05 \cdot 10^{-5}$ GeV$^{-1}$. 

\begin{figure}
\resizebox{0.5\textwidth}{!}{%
  \includegraphics{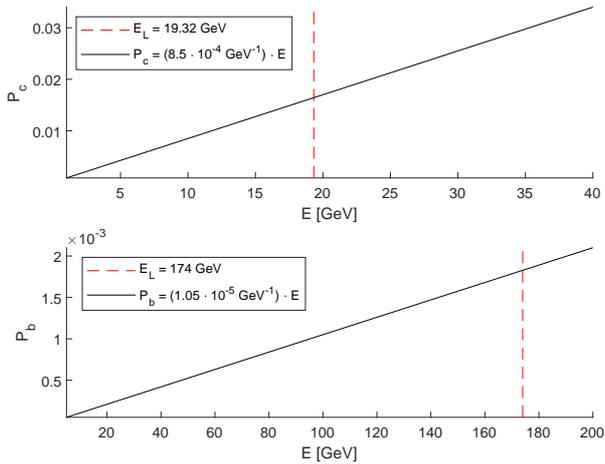}
}
\caption{The energy depenedence of the quark creational probabilities and the limiting energies for the charm (upper panel) and bottom (bottom panel) quarks.}
\label{fig:3}       
\end{figure}

The energy dependent quark creational probabilities and the suppressions can be seen in Fig.\ref{fig:3}, where the limiting energy - the energy, where the suppression from the constrained quark number distribution breaks down -  is also shown. After the limiting energy, $P_{c,b}$ still increases until a full equilibrium is reached with the other type of quarks, however, we do not intend to use the model until that energy. The probabilities are energy dependent, therefore the previously fitted $P_u$, $P_d$, and $P_s$ values have to be changed accordingly, to $P_i' = P_i - P_c/3 - P_b/3$, where $i=u$,$d$,$s$. Due to the smallness of the charm and bottom probabilities, the corrections are very small at the energy range we are interested in, so in practical calculations $P_i' \approx P_i$ can be used.

In the next section the results for different onium state production, as well as some open charm calculations are shown for proton-proton, pion-proton, and proton-antiproton collisions.

\section{Results}
\label{sec:3}
\begin{figure}
\resizebox{0.5\textwidth}{!}{%
  \includegraphics{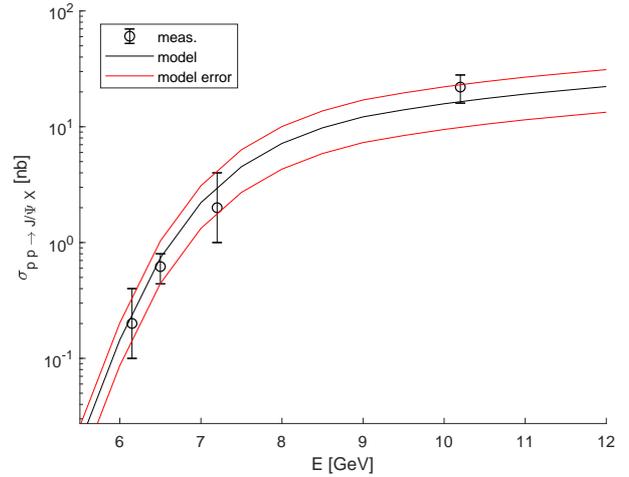}
}
\caption{Model calculation for the $p p \rightarrow J/\Psi X$ inclusive cross section. The black and red lines are the model result and its error respectively, while the circles are measurement points from \cite{30,31,32,33,34,35,36,37,38,39,40,41}.}
\label{fig:5}       
\end{figure}

\begin{figure}
\resizebox{0.5\textwidth}{!}{%
  \includegraphics{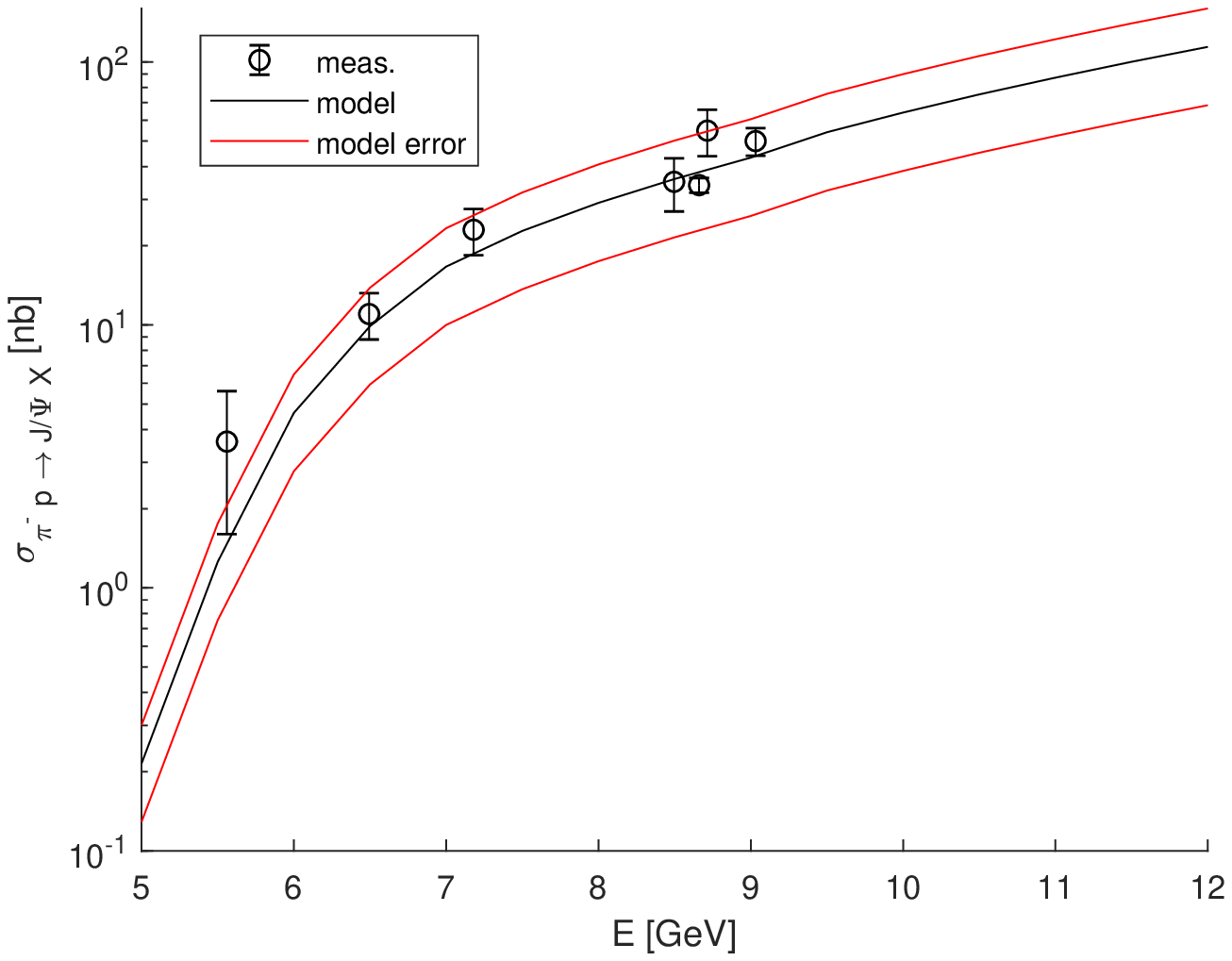}
}
\caption{Model calculation for the $\pi^- p \rightarrow J/\Psi X$ inclusive cross section. The black and red lines are the model result and its error, while the circles are measurement points from \cite{30,31,32,33,34,35,36,37,38,39,40,41}.}
\label{fig:6}       
\end{figure}

\begin{figure}
\resizebox{0.5\textwidth}{!}{%
  \includegraphics{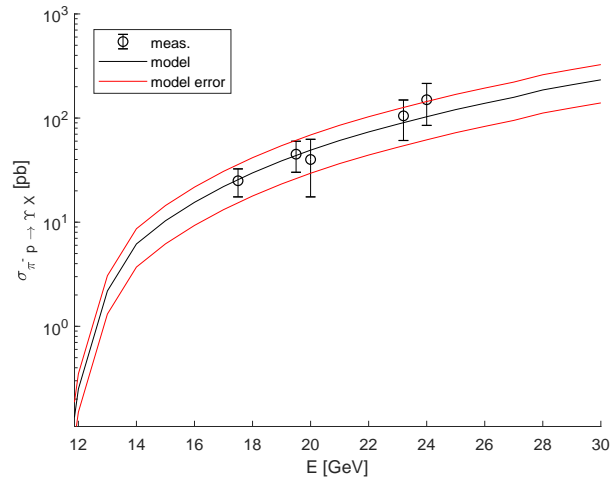}
}
\caption{Model calculation for the $\pi^- p \rightarrow \Upsilon X$ inclusive cross section. The black and red lines are the model result and its error respectively, while the circles are measurement points from \cite{42,43}.}
\label{fig:7}       
\end{figure}

After the quark creational probabilities for heavy quarks are obtained, it is possible to calculate inclusive cross sections for charmed and bottomed hadrons as well. The first reaction shown is the energy dependence of the $p p \rightarrow J/\Psi X$ process, which is used to fit the charm quark creational probability. The decays from the excited $\Psi(3686)$, $\chi_{c1}$, $\chi_{c2}$ states are also included in the calculations. The results can be seen in Fig.\ref{fig:5}, where as expected, the agreement with the data is really good. The second example is the $\pi^- p \rightarrow J/\Psi X$ process, which is expected to give a few times larger cross section compared to the proton-proton case. The results can be seen in Fig.\ref{fig:6}, where again the agreement is still satisfactory. The third example is the $\pi^- p \rightarrow \Upsilon(1S) X$ process, where apart from the direct bottomonium production, the decays from the excited onium states $\Upsilon (2S)$, $\Upsilon (3S)$, $\chi_{b1}(1P)$, $\chi_{b2}(1P)$, $\chi_{b1}(2P)$, and $\chi_{b2}(2P)$ are included. The results can be seen in Fig.\ref{fig:7}. The agreement between measurement and model in the previous three processes are, however expected as these were the ones used to fit the quark creational probabilities. 

\begin{figure}
\resizebox{0.5\textwidth}{!}{%
  \includegraphics{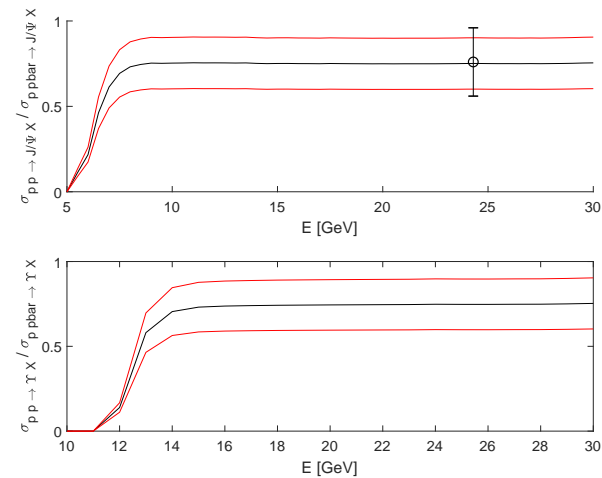} 
}
\caption{Inclusive charmonium (upper panel) and bottomonium (lower panel) cross section suppression ratios in proton-proton to proton-antiproton collisions. The $p \overline{p} \rightarrow J /\Psi$ and $p \overline{p} \rightarrow \Upsilon(1S)$ one body final state Breit-Wigner cross sections are not included in the model calculations. The measured value for charmonium production is taken from \cite{59}.}
\label{fig:9}       
\end{figure}

The fourth process is the $p p \rightarrow \Upsilon X$, where $\Upsilon = \Upsilon(1S) + \Upsilon(2S) + \Upsilon(3S)$. The measured data points are at $\sqrt{s}=29.1$ GeV, with proton beam and five different target nuclei (Be, Al, Cu, Ag, W), taken from \cite{44}. The results can be seen in Fig.\ref{fig:9}, where on the upper side all the five measured data points are shown with the model calculation at $\sqrt{s}=29.1$ GeV, while at the bottom the energy dependence of the model calculation is shown.
\begin{figure}
\resizebox{0.5\textwidth}{!}{%
  \includegraphics{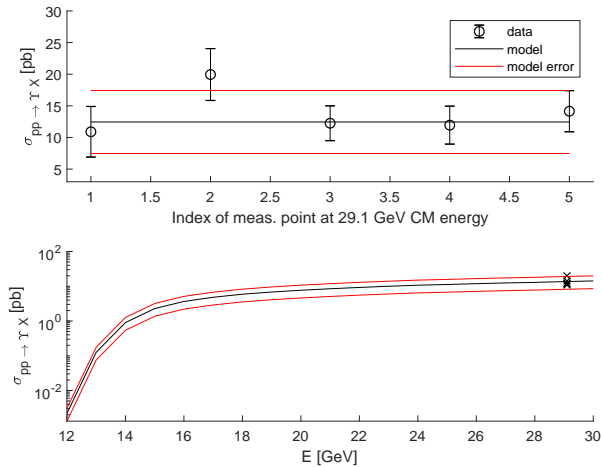}
}
\caption{Model calculation for the $p p \rightarrow \Upsilon X$ inclusive cross section. The black and red lines are the model result and its error, while the circles and crosses are measurement points from \cite{44}.}
\label{fig:8}       
\end{figure}

For the fifth and sixth processes, we checked the proton-proton collision suppression to proton-antiproton collisions in the inclusive charmonium and bottomonium channels. In these calculations, we excluded the one body final states in the proton-antiproton collisions e.g. $p \overline{p} \rightarrow J/\Psi$, which only gives a significant contribution near threshold. The results can be seen in Fig.\ref{fig:9}.
The proton-antiproton collision is therefore favorable in constrast to the proton-proton case, however its production rate is still lower than in pion-proton collisions. This is not just a curiosity on its own, because in heavy ion collisions an interesting problem is the propagation of onium states in dense nuclear matter, as they could have a significant mass shift, which is proportional to the gluon condensate \cite{45,46}, so if one could measure the mass shift of e.g. $\Psi(3686)$ the nuclear expectation value of the gluon condensate at a specific nuclear density can be deduced. Due to the really small cross sections it is necessary to create the most onium states one can get, which is one reason to use antiprotons or pions in these investigations. 

Some of our transport calculations\cite{47,48} requires the determination of the direct charmonium production, without the decays from higher states as well. As there are only just a few measured values for e.g. direct $\Psi (3686)$ production, we checked the ratios defined in Eq.\ref{eq:14} and in Eq.\ref{eq:15} for the $\chi_{c1}$, $\chi_{c2}$ and $\Psi (3686)$ particles:
\begin{equation}
\label{eq:14}
R_{\chi_{c1} + \chi_{c2}}^{AB} = \sum_{i=1}^{2 }\frac{\text{Br}(\chi_{ci} \rightarrow J/\Psi) \sigma_{AB \rightarrow \chi_{ci} X} }{\sigma_{AB \rightarrow J/\Psi X}}
\end{equation}

\begin{equation}
\label{eq:15}
R_{\Psi (3686)}^{AB} = \frac{\text{Br}(\Psi (3686) \rightarrow J/\Psi) \sigma_{AB \rightarrow \Psi (3686) X} }{\sigma_{AB \rightarrow J/\Psi X}},
\end{equation}
where $A$ and $B$ represents the colliding particles, in this case $\pi^-$ and $p$, and Br is the branching fraction of the particles to decay into a $J/\Psi$ meson. The measured ratios are collected from \cite{49,50,51,52,53,54} and the comparison with the model calculations at different collision energies are shown in Table.\ref{table:1} and in Table.\ref{table:2}, where a really good match with the data is achieved. This allows us to calculate the full energy dependence of the direct production of these states and put it into our transport simulations.

\begin{figure}
\resizebox{0.5\textwidth}{!}{%
  \includegraphics{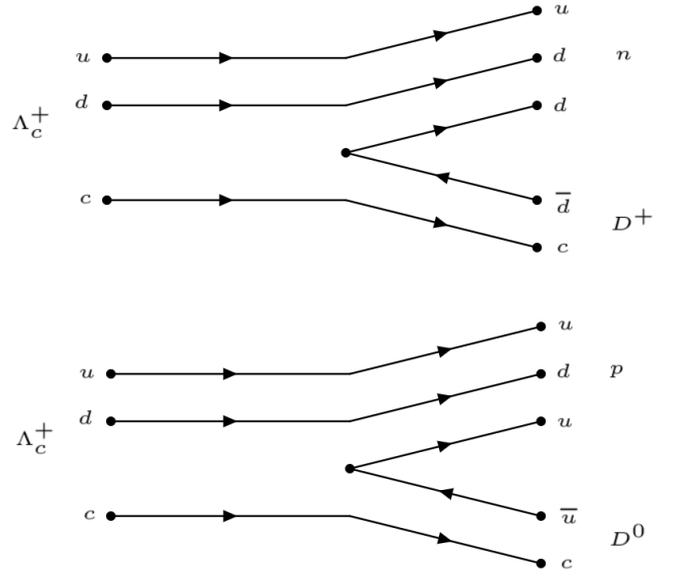} 
}
\caption{Quark flow diagrams to the decays $\Lambda_c^+ \rightarrow D^+ n$ and $\Lambda_c^+ \rightarrow D^0 p$.}
\label{fig:10}       
\end{figure}

\begin{figure}
\resizebox{0.5\textwidth}{!}{%
  \includegraphics{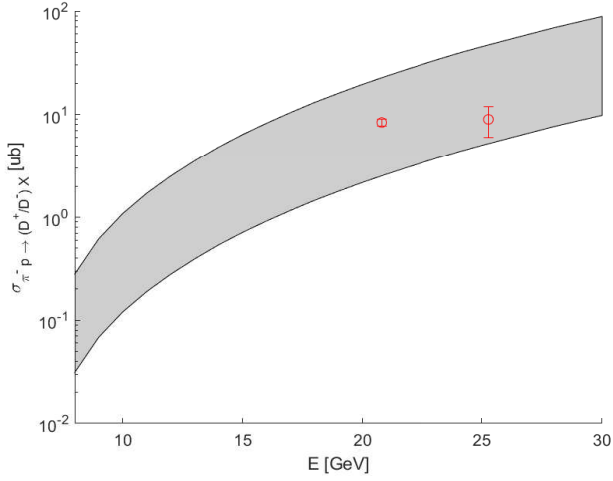}
}
\caption{Estimated cross section interval for the $\pi^- p \rightarrow D^+/D^-$ process, with varied branching fraction asymetries. The data is taken from \cite{56,57,58}.}
\label{fig:11}       
\end{figure}

\begin{figure}
\resizebox{0.5\textwidth}{!}{%
  \includegraphics{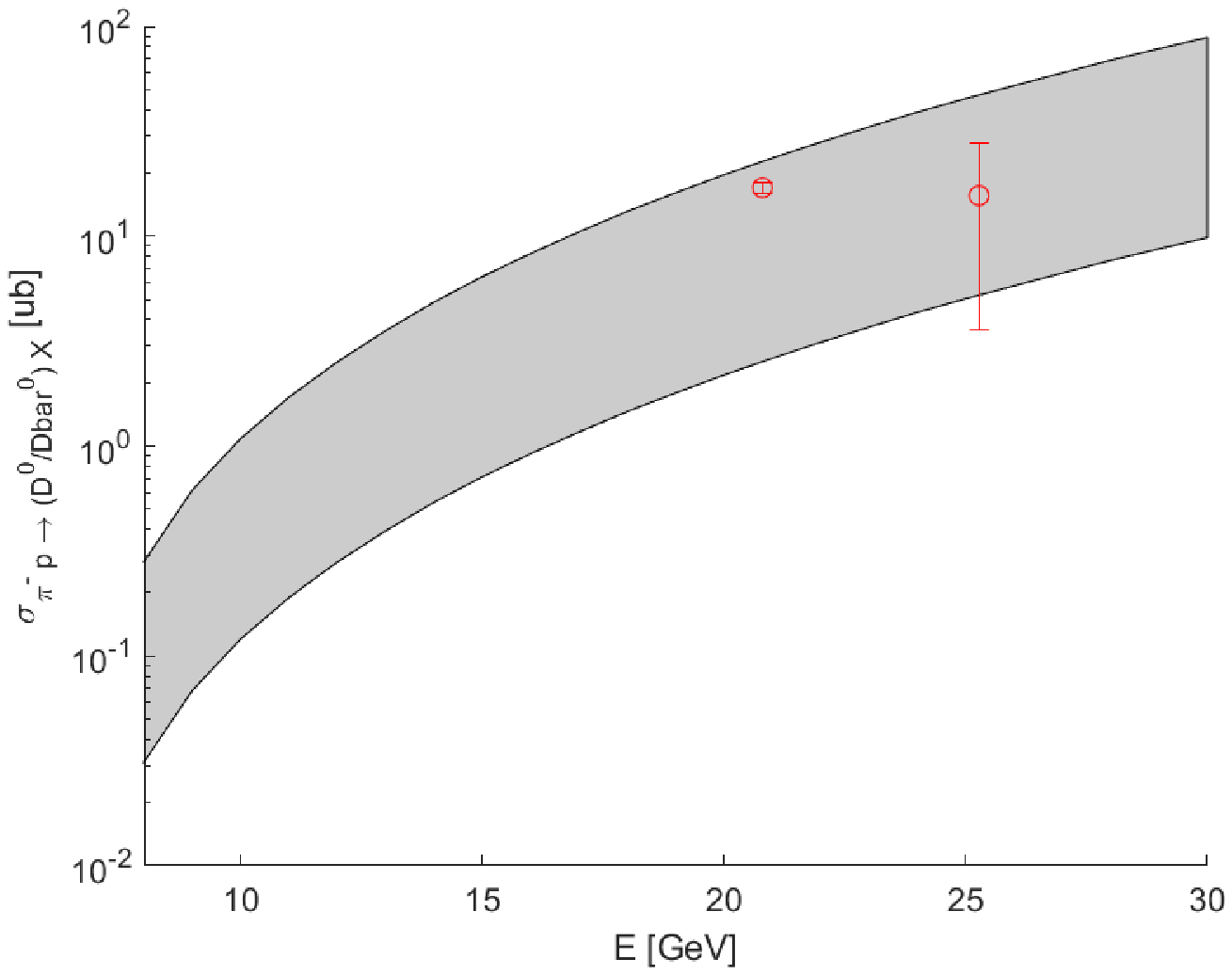}
}
\caption{Estimated cross section interval for the $\pi^- p \rightarrow D^0/\overline{D}^0$ process, with varied branching fraction asymetries. The data is taken from \cite{56,57,58}.}
\label{fig:12}       
\end{figure}

\begin{figure}
\resizebox{0.5\textwidth}{!}{%
  \includegraphics{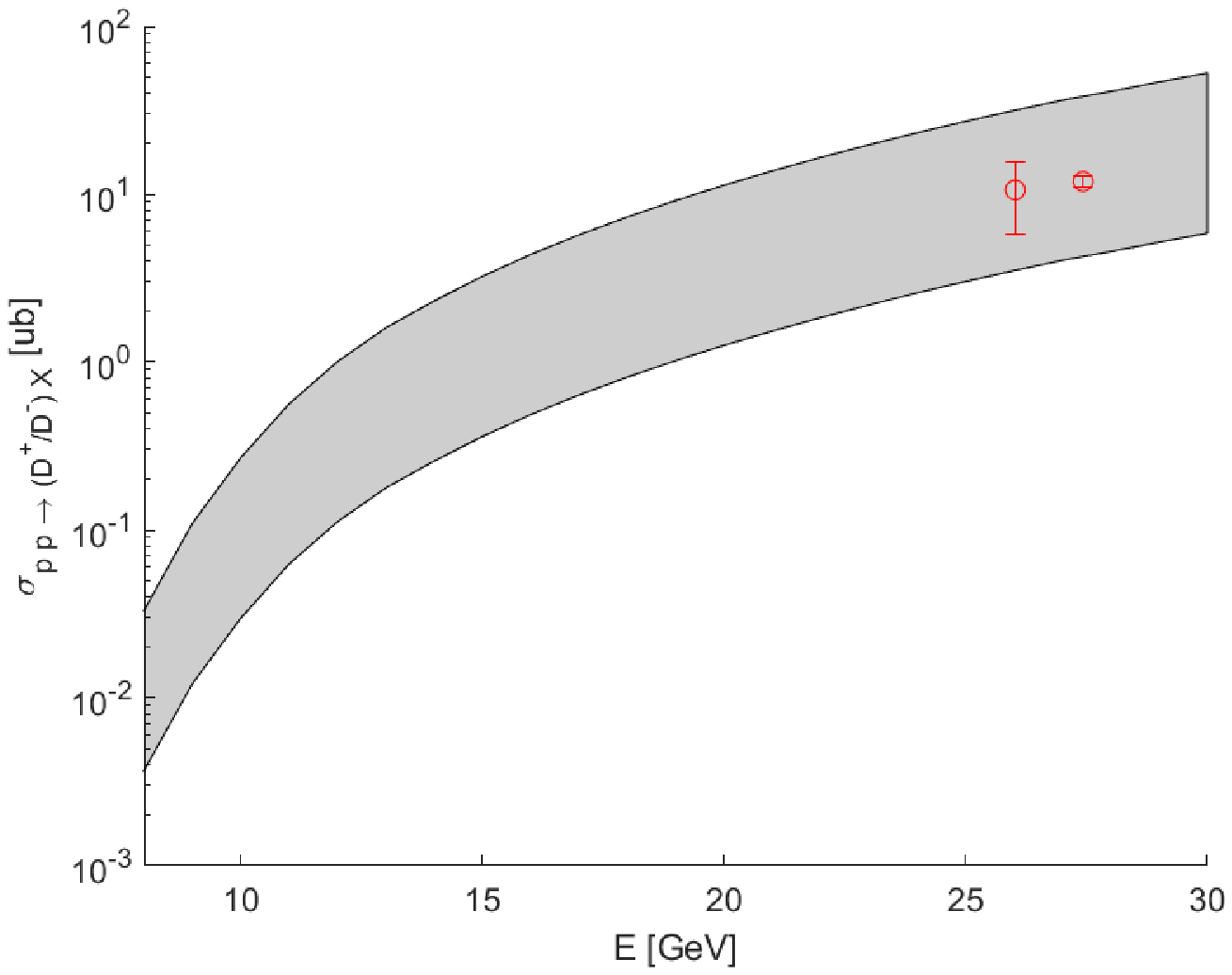}
}
\caption{Estimated cross section interval for the $p p \rightarrow D^+/D^-$ process, with varied branching fraction asymetries. The data is taken from \cite{56,57,58}.}
\label{fig:13}       
\end{figure}

\begin{figure}
\resizebox{0.5\textwidth}{!}{%
  \includegraphics{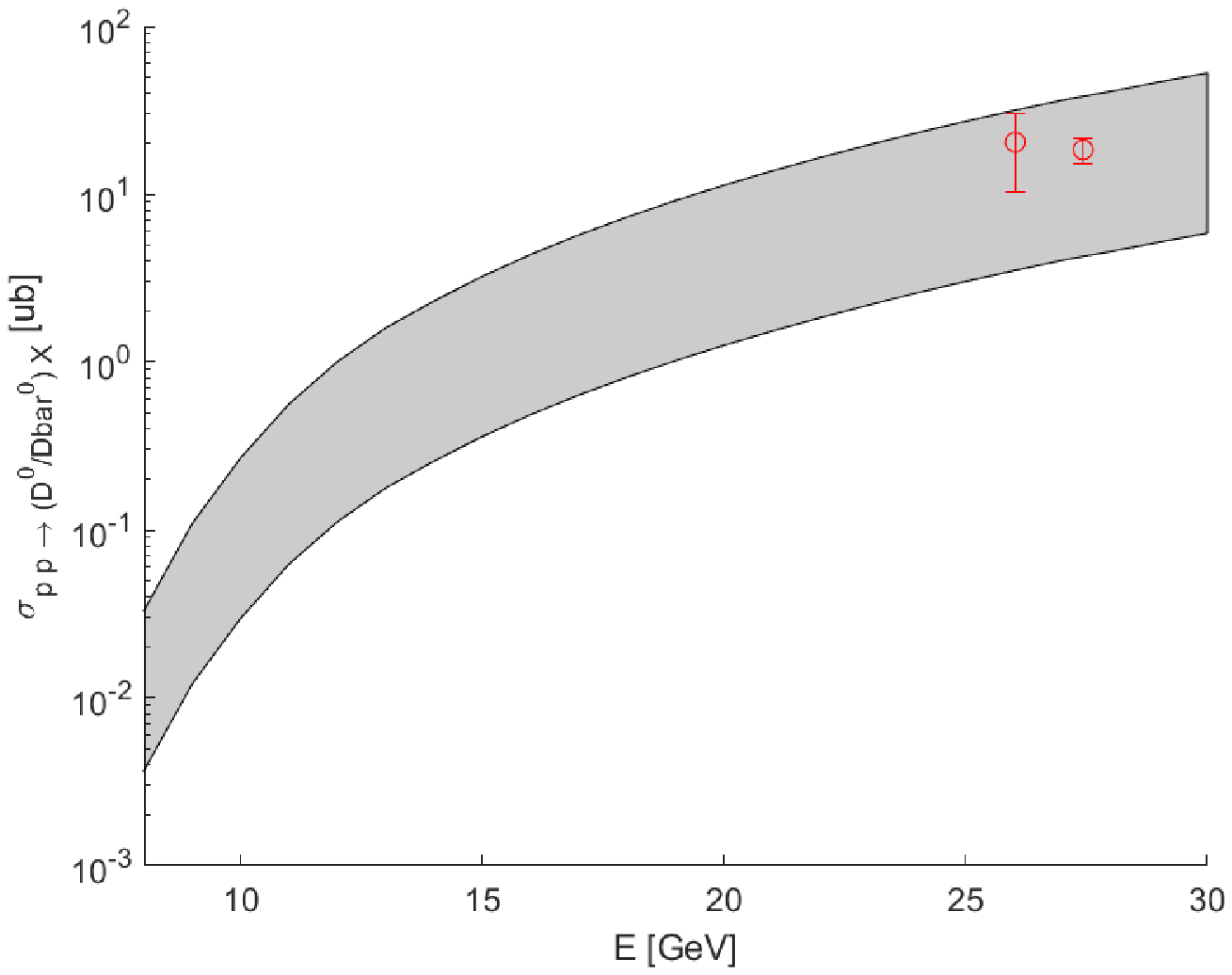}
}
\caption{Estimated cross section interval for the $p p \rightarrow D^0/\overline{D}^0$ process, with varied branching fraction asymetries. The data is taken from \cite{56,57,58}.}
\label{fig:14}       
\end{figure}

\begin{table}[]
\centering
\begin{tabular}{|c|c|ccc|}
\hline
                       &                       & \multicolumn{3}{c|}{}    \\
\multicolumn{1}{|l|}{} & \scalebox{1.2}{$R_{\Psi (2S)}^{pN}$}                    & \multicolumn{3}{c|}{ \scalebox{1.2}{$R_{\Psi (2S)}^{\pi N}$}  }  \\
\multicolumn{1}{|l|}{} & \multicolumn{1}{l|}{} & \multicolumn{3}{l|}{}    \\ \hline
$\sqrt s$ [GeV]                      & $23.7$                  & $18.6$  & $20.5$  & $23.74$    \\
meas. [\%]                   & $5.5\pm1.6$              & $8$     & $11\pm5$ & $7.6\pm2.3$ \\
model [\%]                 & $6.2\pm2.4$                 & $8.5\pm3.4$ & $8\pm3.2$   & $7.2\pm2.8$    \\ \hline
\end{tabular}
\caption{Total $J/\Psi$ production ratio to the direct $\Psi(3686)$ production cross section. The measured values are collected from \cite{49,50,51,52,53,54}.}
\label{table:1}
\end{table}
\begin{table}[]
\centering
\begin{tabular}{|c|ccc|}
\hline
      &           &           &           \\
      &           & \scalebox{1.2}{$R_{\chi_{c1} + \chi_{c2}}^{\pi N}$}        &           \\
      &           &           &           \\ \hline
$\sqrt s$ [GeV]       & $18.6$      & $18.9$      & $23.74$      \\
meas. [\%]   & $30.5\pm6.5$ & $31\pm14$    & $34\pm4$     \\
model [\%]   & $35.7\pm7.1$ & $35.9\pm7.1$ & $36.4\pm7.2$ \\ \hline
\end{tabular}
\caption{Total $J/\Psi$ production ratio to the direct $\chi_{c1}$, $\chi_{c2}$ production cross sections. The measured values are collected from \cite{49,50,51,52,53,54}.}
\label{table:2}
\end{table}
The following few processes are somewhat more complicated than the previous ones. These are the inclusive open charm productions in $\pi^- p$ and in $p p$ collisions. The main problem is that our model is heavily based on the known resonances, excited states, which could decay into hadrons we are interested in, and if the branching fractions or the quantum numbers of such particles are not well known, then it could not be included in the calculations. The problem with open charm (and open bottom) cross sections is therefore lies in the fact, that there are many excited mesons and baryons that could decay into ground state open charm/bottom mesons, where the branching fractions are not known at all. Nevertheless, we still try to give an order of magnitude estimation to these cross sections, namely to: $\pi p \rightarrow D^{+}/D^- X$, $\pi p \rightarrow D^{0}/\overline{D}^0 X$, and the $p p \rightarrow D^{+}/D^- X$, $p p \rightarrow D^{0}/\overline{D}^0 X$. To all of these processes the resonances $D^*(2007)^0$, $D^*(2010)^{\pm}$, $D^*(2300)^0$, $D^*(2300)^{\pm}$, $D_1(2420)^0$, $D_2^*(2460)^{0}$, $D_2^*(2460)^{\pm}$, $D_3^*(2750)$, $\Lambda_c(2860)^+$, $\Lambda_c(2880)^+$, $\Lambda_c(2940)^+$ are included, whose quantum numbers are at least fairly well known in the PDG. However, the branching fractions are not known at all, and in the excited $\Lambda_c$ baryon resonances even less is known than the $D$ mesons. To be able to give an estimation, we introduced an asymmetry parameter $r$ defined as:
\begin{equation}
\label{eq:16}
r = \frac{\text{Br}(X \rightarrow D^+/D^- )}{\text{Br}(X \rightarrow D^0/\overline{D}^0)},         
\end{equation}
with the assumption that every other decay is negligible, so $\text{Br}(X \rightarrow D^+/D^- ) + \text{Br}(X \rightarrow D^0/\overline{D}^0) \approx 1$ . The ratio thus gives the fraction that a resonance '$X$' e.g. $D^*(2300)^0$ giving a $D^+$ to the expense of $D^0$, and varied it between $[0.11,9]$, which for the branching ratios means Br$_i \in [0.1,0.9]$. This applies to the differently charged and neutral resonances with different possible final states as well. This is reasonable as for example in the $D^{*}(2007)^+$ decays, where the branching fractions are known, the asymmetry in the decay products is $r\approx 0.46$. For the $\Lambda$ resonances, where the final state products are also not well known, we assumed it could decay as $\Lambda_c^+ \rightarrow p D^0$, $\Lambda_c^+ \rightarrow n D^+$, and for the antiparticles $\overline{\Lambda}_c \rightarrow \overline{p} \overline{D}^0$, $\overline{\Lambda}_c \rightarrow \overline{n} D^-$ according to the quark flow diagrams in Fig.\ref{fig:10}, with the same asymmetry variations as in the meson resonances. 
In \cite{55} an effective model calculation estimated the asymmetry to $r\approx 1$, and it is highly possible that each resonance has different branching fractions, so the best we can do at this point is to stay at the varied asymetries and give an order of magnitude estimation. It could be an interesting further step to include all the possible resonances and branching ratios from e.g. effective model calculations. Using the mentioned assumptions, the results are shown in Fig.\ref{fig:11}-Fig.\ref{fig:14}, where only cross section intervals are shown due to the varied asymmetries. It can be seen that an order of magnitude estimation is very much achievable, even with the high uncertainty in the parameters.

\section{Conclusions}
\label{sec:4}
We extended our model with the inclusion of charmed and bottomed hadrons and estimated the charm and bottom quark creational probabilities, assuming a simple linear relationship between the probability and the energy $P_{c,b}=a_{c,b}E$. A fit is made for the two slope parameters $a_c$, and $a_b$, using inclusive charmonium and bottomonium data in proton-proton and in pion-proton collisions. The suppression of the higher charmonium states $\Psi(3686)$, $\chi_{c1}$, and $\chi_{c2}$ to the direct $J/\Psi$ production is also calculated and compared to the measured ratios in $\pi^- p$ and $p p$ collisions, giving a really good match with the data. The model is further validated through the processes $p p \rightarrow J/\Psi X$, $\pi^- p \rightarrow J/\Psi X$, $p p \rightarrow \Upsilon X$, and $\pi^- p \rightarrow \Upsilon X$. Estimations were made to proton-antiproton collisions and the ratios $\sigma_{p\overline{p} \rightarrow J/\Psi X}/\sigma_{p p \rightarrow J/\Psi X}$, $\sigma_{p \overline{p} \rightarrow \Upsilon X}/\sigma_{p p \rightarrow \Upsilon X}$ are calculated.  It can be concluded that at low energies proton-antiproton and pion-proton collisions are very much favorable if one wants to produce charmonium particles. For further validation open charm production cross sections were also estimated, namely the $\pi^- p \rightarrow D^+/D^- X$, $p p \rightarrow D^+/D^- X$, $\pi^- p \rightarrow D^0/\overline{D}^0 X$, and the $p p \rightarrow D^0/\overline{D}^0 X$ processes. The main problem with these calculations are the missing branching fractions in the PDG, which are needed to make reliable estimates, however introducing an asymmetry parameter, an order of magnitude estimation was made, which is in agreement with the measurements. With these extensions the model is now considered complete regarding non-exotic states. It is however a further question how to include e.g. tetra-quarks into the model, as there are no measurements for these states at low energies, where the model works.

\bibliographystyle{unsrt}
 \bibliography{Main_Document}

\end{document}